\documentstyle[times,epsfig]{jaa}

\def\subF #1{#1_{\lower2pt\hbox{$\scriptstyle{\rm F}$}}}

\begin{document}
\title[Mass radius scaling near the Chandrasekhar Limit]{Mass radius scaling near the Chandrasekhar Limit}
\author[Sayan Chakraborti]{Sayan Chakraborti\thanks{E-mail:
sayan@tifr.res.in}\\
Tata Institute of Fundamental Research (TIFR), Mumbai 400-005, India}
\pubyear{0000}
\volume{00}


\maketitle

\label{firstpage}

\begin{abstract}
The mass radius relationship of white dwarfs, near the Chandrasekhar Limit, is derived for a toy model of uniform density, using the variational principle. A power law scaling, reminiscent of those found in 2nd order phase transitions, is obtained. The derived exponent is shown to explain the relationship obtained by numerically integrating the TOV equations with the equation of state for a relativistic Fermi gas of electrons.
\end{abstract}


\section{Introduction}

A white dwarf can be sufficiently described by special relativistic physics without taking account of general relativity as the internal energy and pressure are both very much less than the rest mass density. Following Weinberg (1972) we use simple hydrostatics to get the equations of stellar equilibrium.
\begin{eqnarray}
   &&\frac{dP}{dr}=-\frac{GM(r)\rho(r)}{r^2}\;, 
   \quad P(r\!=\!0)\equiv P_{\rm c} \;; 
 \label{HydroEqsa} \\
   &&\frac{dM}{dr}=+4\pi r^{2}\rho(r) \;, 
   \hspace{0.32in} M(r\!=\!0)\equiv 0 \;, 
 \label{HydroEqsb}
\end{eqnarray}
Thus, given an equation of state $p = p(\rho)$ we can obtain the mass radius relationship of the given object by integrating up to the surface indicated by $p=0$.

\section{Objectives}
Assuming cold, spherically symmetric, non-rotating white dwarfs in hydrostatic equilibrium, the equation of state is provided by the pressure of the Fermi gas of electrons. The equations can be integrated analytically if the equations of state are polytropic, so the non-relativistic limit gives us a mass radius relationship of $R \sim M^{-1/3}$, while the relativistic limit gives a mass, the Chandrasekhar Limit, independent of the radius. However the nature of the mass radius relationship in the vicinity of the critical mass is not revealed in the polytropic analysis. The objective of this work is to evaluate the radius, of a toy model proposed by Jackson et al (2005) near its critical mass. Its radius is shown to have a power law dependence on the mass. This scaling relation is used to estimate the mass radius relationship of a white dwarf near the Chandrasekhar Limit. The critical exponent evaluated in the present work is shown to fit very well with simulations for white dwarfs with a relativistic Fermi gas.

\section{Uniform Density Model}

Jackson et al (2005) define an uniform density star by
\begin{equation}
  \rho(r)=
  \begin{cases}
      \rho_{0}=3M/4\pi R^{3}\;,  {\rm if} \quad r\le R\;; \\
       0\;,                      {\rm if} \quad r>R   \;,
  \end{cases}
 \label{UniformDensity}
\end{equation}
and use elegant scaling arguments to transform to dimensionless coordinates defined by
\begin{equation}
  \overline{M}\!\equiv\!M/M_{0} 
  \quad{\rm and}\quad
  \overline{R}\!\equiv\!R/R_{0} \;.
 \label{BarQuantities}
\end{equation}
where $M_{0} = 2.650\,M_{\odot}$ and $R_{0}=8\,623~{\rm km}$, so that the equations of stellar structure become simply,
 \begin{eqnarray}
   &&\frac{d\subF{x}}{d\overline{r}}=
    f(\overline{r};\subF{x},\overline{M}) \;, 
    \quad \subF{x}(\overline{r}\!=\!0)\equiv
    x_{\lower2pt\hbox{$\scriptstyle{\rm Fc}$}}\;; \\
   &&\frac{d\overline{M}}{d\overline{r}}=
      g(\overline{r};\subF{x},\overline{M}) \;,
    \quad\overline{M}(\overline{r}\!=\!0)\equiv 0 \;,
 \end{eqnarray} 
 \label{HydroEqsScaled}
where the two functions on the right-hand side of the equations
($f$ and $g$) are given by
\begin{eqnarray}
     &&f(\overline{r};\subF{x},\overline{M})\equiv 
    -\frac{5}{3}
     \frac{\overline{M}}{\overline{r}^{2}} 
     \frac{\sqrt{1+\subF{x}^{2}}}{\subF{x}}\;; \\
     &&g(\overline{r};\subF{x},\overline{M})\equiv
    +3\,\overline{r}^{2}\subF{x}^{3} \;.
 \label{fandg}
\end{eqnarray}
Where $\subF{x}\!=\!\hbar\subF{k}c/mc^{2}$ is the ration of the kinetic energy and rest mass of the electrons. Taking advantage of the scaling relations, they express the energy per electron in units of the electron rest energy as
%
\begin{equation}
  f_{total}(\overline{M},x_{\rm F})
  =-\overline{M}^{2/3}x_{\rm F}+3\pi^{2}
  \frac{{\overline{\cal E}}(\subF{x})}{\subF{x}^{3}} \;.
 \label{fTotal}
\end{equation}
Where the first term is the contribution from gravitation and the second term may be evaluated from Fermi-Dirac statistics as
\begin{eqnarray}
 && \nonumber \overline{{\cal E}}(\subF{x})=
  \frac{1}{8\pi^{2}}\Big[\subF{x}\left(1+2\subF{x}^{2}\right)
  \sqrt{1+\subF{x}^{2}}\\
  &&-\ln\left(\subF{x}+
  \sqrt{1+\subF{x}^{2}}\,\right)\Big]\;.
 \label{EBar}
\end{eqnarray}
Hence, the mass radius relation of the star is obtained by 
demanding hydrostatic equilibrium
\begin{equation}
 \left(\frac{\partial f_{total}(\overline{M},x_{\rm F})}
            {\partial x_{\rm F}}\right)_{\overline{M}}=0\;.
\end{equation}

\section{Critical Exponent}
To evaluate a tentative value of the Chandrasekhar Limit, let us evaluate the equation of hydrostatic equilibrium in the limit of $\subF{x}>>1$.
\begin{eqnarray}
\left(\frac{\partial f_{total}(\overline{M},x_{\rm F})}
            {\partial x_{\rm F}}\right)_{\overline{M}}=-\overline{M}^{2/3}+3\pi^{2}\left(\frac{\partial}
            {\partial x_{\rm F}}
  \frac{{\overline{\cal E}}(\subF{x})}{\subF{x}^{3}}\right)_{\overline{M}}
\end{eqnarray}
Substituting ${\overline{\cal E}}(\subF{x})$, neglecting logarithmic terms with respect to linear terms and binomially expanding the square root up to terms of order $\subF{x}^-2$, we have
\begin{equation}
-\overline{M}^{2/3}+\frac{3}{4}-\frac{3}{4 \subF{x}^2} = 0
\end{equation}
By the choice of the dimensionless coordinates, cumbersome constants in front of equations have been set to $1$. So, substituting
\begin{equation}
\subF{x}^3 = \frac{\overline{M}}{\overline{R}^3}
\end{equation}
we get
\begin{equation}
\overline{R}^2 = \overline{M}^{2/3}-\frac{4}{3}\overline{M}^{4/3}
\end{equation}
This tends to zero as $\overline{M}$ tends to $\overline{M}_{cr}=(3/4)^{3/2}$. Taylor expanding about this point, we have
\begin{equation}
\overline{R} \sim \Big[1-\frac{\overline{M}}{\overline{M}_{cr}}\Big]^{1/2}
\end{equation}
It can be immediately seen from the above relation that the order parameter $\overline{R}$ has a critical exponent $\beta=1/2$.

\begin{figure*}
\psfig{file=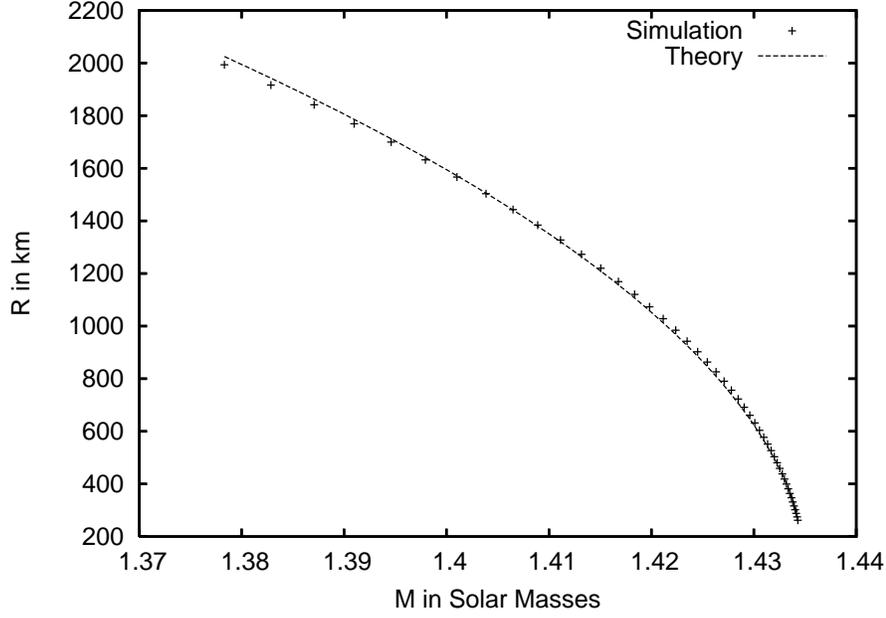,angle=-90,width=12cm}
\caption{Mass radius relationship}
\end{figure*}

\section{Numerical Analysis}
The equations of structure are now integrated numerically, using a $10^6$ points adaptive mesh, from the central region until the pressure falls off to zero. This process is carried on for different central densities in the region $10 < \subF{x} <100$ to trace out the mass radius relationship near the Chandrasekhar Limit.

\section{Results}
Using $\beta=1/2$, the radius is written in terms of the mass as
\begin{equation}
R = R_{constant} \Big[1-\frac{M}{M_{cr}}\Big]^{1/2}
\label{final}
\end{equation}
It is best-fitted with the simulated mass radius data to give a best fit of $R_{constant}=10149 \pm 25$ kms and $M_{cr}=1.43544 \pm 0.00004$ Solar masses, for a Helium white dwarf. Investigating the parameter space, the critical exponent $\beta$ is constrained to $0.499 \pm 0.005$, which matches well with the prediction. Hence Eq. \ref{final} gives us the mass radius relationship of white dwarfs near the Chandrasekhar Limit.

\section*{Acknowledgments}
I would like to thank Mr. R. Loganayagam and Ms. Mehuli Mondal for illuminating discussions.

\label{lastpage}

\end{document}